\documentclass[prb,preprint]{revtex4}

\pdfoutput=1

\usepackage{graphicx}

\begin{document}

\title{Superlattice origin of incommensurable density waves in $\mathbf{La_{2-x}Ae_{x}CuO_{4}}$ $\mathbf{(Ae = Ba, Sr)}$

\medskip }

\date{July 27, 2013} \bigskip

\author{Manfred Bucher \\}
\affiliation{\text{\textnormal{Physics Department, California State University,}} \textnormal{Fresno,}
\textnormal{Fresno, California 93740-8031} \\}

\begin{abstract}
In line with the Coulomb-oscillator model of superconductivity, loop currents of excited $3s$ electrons from $O^{2-}$ ions, passing in the $CuO_{2}$ plane through nuclei of nearest-neighbor oxygen quartets, create the antiferromagnetic phase of undoped copper oxides. Holes, introduced by alkaline-earth doping of $La_{2}CuO_{4}$, destroy the loop currents, thereby weakening antiferromagnetism until it disappears at doping $x = 0.02$. Further doping of $La_{2-x}Ae_{x}CuO_{4}$ gives rise to incommensurate free-hole density waves whose wavelength is determined by the spacing of a doping superlattice. Modulating the ordering of the ions' magnetic moments, the charge-density wave, of incommensurability $2 \delta$, causes a magnetic density wave of incommensurability $\delta$. The formula derived for $\delta (x)$ is in excellent agreement with data from X-ray diffraction and neutron scattering.

\end{abstract}

\maketitle


\section{INTRODUCTION}

By the semiclassical Coulomb-oscillator model of superconductivity,\cite{1} non-resistive current is carried by axial Coulomb oscillations of cystal atoms' outer $s$ electrons through neighbor nuclei if their accompanying lateral oscillation is sufficiently confined to prevent lateral overswing. It is widely believed that high-temperature superconductivity of copper oxides occurs in their $CuO_{2}$ planes. While free copper ions and $O^{2-}$ ions in alkaline-earth oxides possess no outer $s$ electrons---free $O^{2-}$ ions are unstable---compression of the $CuO_{2}$ planes by strong Madelung forces causes an electron reconfiguration of $O^{2-}$ from $2p^{6}$ to $2p_a^1p_b^1p_c^23{s^2}$. Thus, by reducing the occupancy of the $2p_{a}$ and $2p_{b}$ orbitals that are aligned along the $O$-$Cu$ axes, excited $3s$ electrons are created.  They give rise to a variety of phenomena.

The first phenomenon, occuring already in undoped cuprates, is antiferromagnetism.  Other phenomena emerge in conjunction with doping by cation substitution. One can distinguish between ionic and electronic consequences of doping.  In the first case, dopant ions give rise, via electrostatic forces, to ion displacements in the $CuO_{2}$ plane that exert \emph{extra squeeze} of nearby $O^{2-}$ ions. This causes confinement of lateral $3s$ electron oscillation and consequent superconductivity, treated in a previous paper.\cite{2} One of the electronic aspects that accompanies doping---here $p$ doping by substitution of $La^{3+}$ host cations by alkaline-earth ions $Ba^{2+}$ or $Sr^{2+}$---is the occurence of electron deficiency (holes).  This in turn affects antiferromagnetism and gives rise to waves of both charge and magnetic density---topics of the present paper.  The treatment of antiferromagnetism is qualitative, but conceptually necessary for the quantitative treatment of incommensurable density waves.

\section{ANTIFERROMAGNETISM}

Let's first imagine a scenario of $3s$ electrons in an undoped copper-oxide crystal \emph{without} the possibility of antiferromgnetic ordering. The excited $3s$ electrons of squeezed $O^{2-}$ ions would perform Coulomb oscillations through the nuclei of neighbor $O^{2-}$ ions, but also scatter due to lateral overswing and thereby diffuse throughout the $CuO_{2}$ plane. Such diffusion scenarios are schematically depicted in Fig. 1 for two $3s$ electrons---for sake of demonstration, one only by Coulomb oscillations between nearest-neighbor ($nn$) oxygen pairs, the other between next-$nn$ pairs. More realistically, combinations of both cases would occur. The 

\pagebreak
\includegraphics[width=6in]{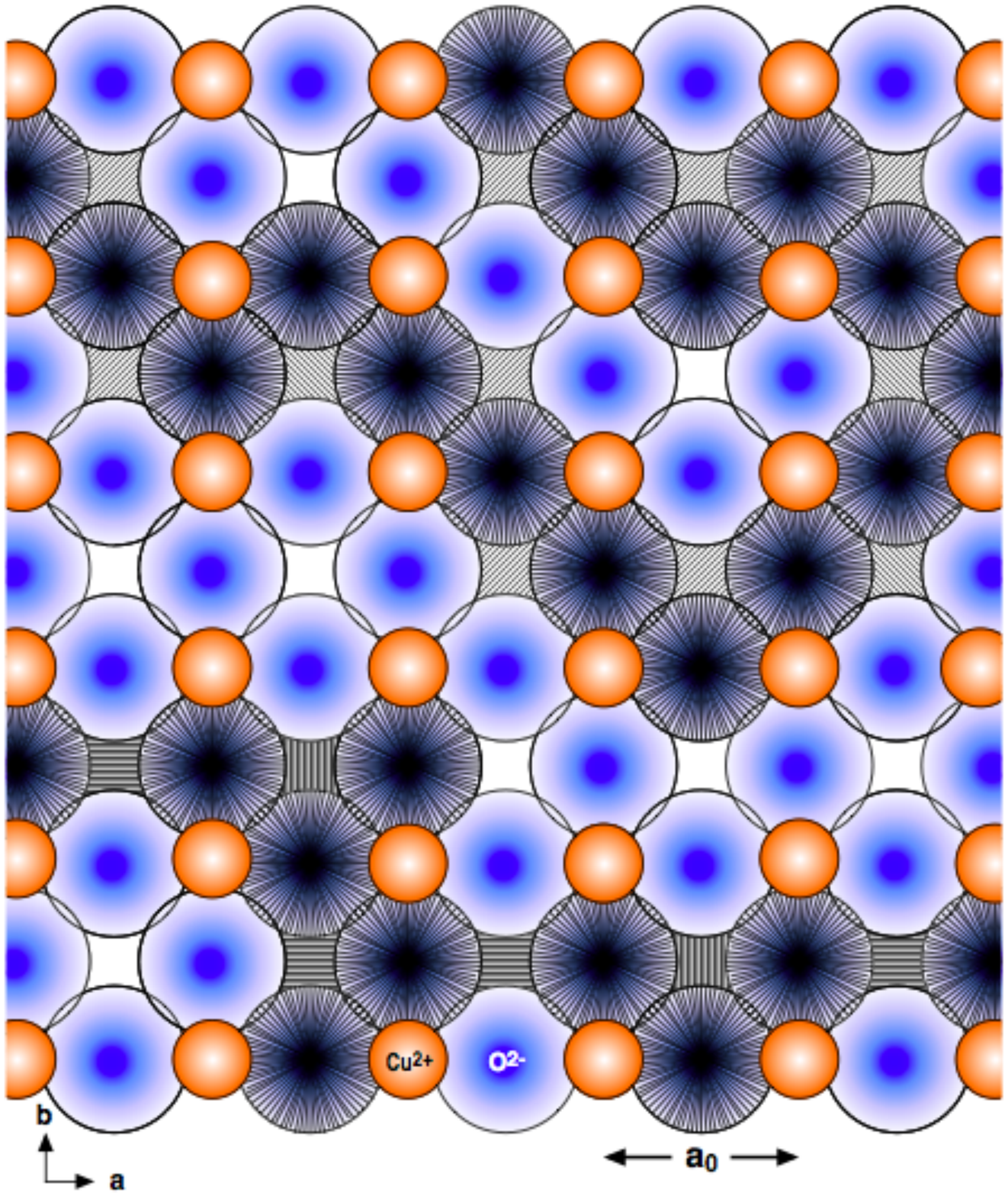}

\noindent FIG. 1. Fictitious time sequence of two diffusing $3s$ electrons in the $CuO_{2}$ plane of a cuprate due to  axial and lateral overswing between $nn$ oxygen sites (top) and between $nnn$ oxygen sites (bottom) \emph{if no antiferromagnetic ordering were possible}. Straight-line hatching indicates $3s$ electron oscillations through oxygen nuclei and across interionic regions.
\pagebreak

\noindent conductivity of the $CuO_{2}$ plane would then be \emph{normal}, not superconductive.

Diffusion of $3s$ electrons in the $CuO_{2}$ plane would cease, however, if a quantum state of lower energy could be reached with \emph{confined} lateral oscillation. Qualitatively, a gain of magnetic interaction energy would result from electron loop currents in the $CuO_{2}$ plane with neighbor currents of opposite circulation, amounting to antiferromagnetic order.  As has been pointed out, at least three atoms (ions) are necessary to anchor a loop current.\cite{3}  For the $CuO_{2}$ plane, diamond-shaped loop currents through \emph{four} $nn$ oxygen nuclei seem appropriate, involving the same concept of axial Coulomb oscillation that also accounts for the emergence of superconductivity in \emph{doped} copper oxides.\cite{2} There are two possibilities of arranging such diamond-shaped loop currents: either around the $Cu^{2+}$ ions or around the empty center of $nn$ $O^{2-}$ quartets, that is, displaced by $(\frac{1}{2},\frac{1}{2},0)\, {a_0}$ from each $Cu^{2+}$ ion.  More information is necessary to decide which case is favorable.  Considering that the  $O^{2-}$ ions in their $2p_a^1p_b^1p_c^23{s^2}$ configuration and the $Cu^{2+}$ ions both have magnetic dipole moments due to uncompensated orbital angular momentum and spin, we want to assume the first case, illustrated in Fig. 2.  The loop currents are depicted by diamond-shaped hatched traces, indicating the laterally confined $3s$ electron flow. Engagement in loop-current circulation prevents the $3s$ electrons from diffusing in the $CuO_{2}$ plane---a qualitative explanation for the antiferromagnetic \emph{insulator}.

The loop currents give rise to magnetic dipole moments at the $Cu^{2+}$ positions, alternating along or against the $+c$ directions according to the $3s$ electron circulation.  It can be assumed that the local magnetic field from the loop currents in turn aligns the magnetic moments of the anchoring four $nn$ $O^{2-}$ ions and of the surrounded  $Cu^{2+}$ ions. The antiferromagnetic order illustrated in Fig. 2 shows a diagonal pattern of loop-current circulation with respect to the planar crystal axes, repeating after two lattice constants, $2a_{0}$, along both the $a$ and $b$ axis.  This corresponds to a reciprocal-lattice point $\textbf{G} = (\frac{1}{2},\frac{1}{2},0)$ in reciprocal-lattice units (r.l.u.). A Bragg peak at  $\textbf{G}$ in neutron scattering provides evidence of antiferromagnetic order in copper oxides.

\pagebreak
\includegraphics[width=6in]{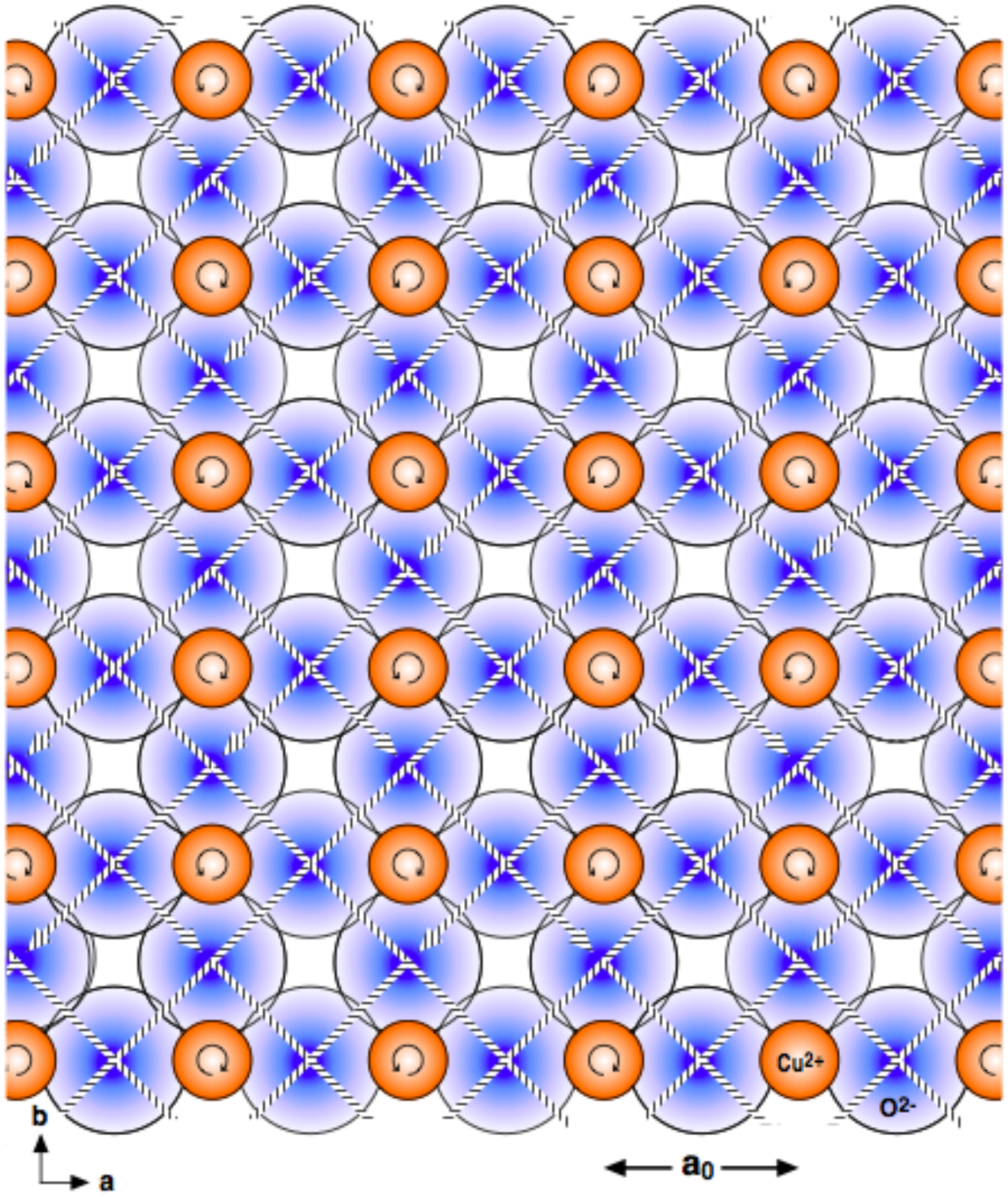}

\noindent FIG. 2. Antiferromagnetic order in the $CuO_{2}$ plane in \emph{undoped} copper oxide due to laterally confined $3s$ electron loop currents (hatched diamond traces) through quartets of nearest-neighbor oxygen nuclei. Their electron-flow circulation is more clearly indicated at the surrounded $Cu^{2+}$ ions.
\pagebreak

\section{DENSITY WAVES}

Doping of $La_{2}CuO_{4}$ with alkaline-earth $Ae$ = $Ba$ or $Sr$, denoted $La_{2-x}Ae_{x}CuO_{4}$, gives rise in each $CuO_{2}$ plane to a planar superlattice of spacing
\begin{equation}
A(x) = \sqrt {\frac{2}{x}} \; {a_0} \; ,
\end{equation}
\noindent where \emph{pairs} of $Ae^{2+}$ ions reside in the layers above and beneath.\cite{2}

In contrast to trivalent lanthanum of the host lattice, each dopant $Ae$ atom provides, upon ionization, only \emph{two} instead of three electrons to the ionization of $O$ atoms. With respect to the undoped crystal, each deficient electron is regarded as a ``hole'' of charge $+e_{0}$, hence the term ``hole doping.'' By Eq. (1) a doping ratio $x$ introduces a \emph{pair} of holes per planar superlattice domain, that is, a hole density per planar unit area (or per $Cu^{2+}$ ion),
\begin{equation}
p(x) = \frac{{2a_0^2}}{{A{{(x)}^2}}} \; .
\end{equation}

How do such holes affect the antiferromagnetic phase? In order to see the basic concept, pick an $O^{2-}$ ion in Fig. 2 (not at the margin) and assume that a hole has settled there, reducing the ion to $O^{-}$. This spoils the two diamond-shaped $3s$ loop currents that pass through that oxygen nucleus, along with the local antiferromagnetism created by both loop currents. When the hole moves on from the oxygen site under consideration, loop currents can resume and antiferromagnetism will be locally restored. As the scenario illustrates, two dynamic processes compete: (1) Diffusion of the loop-current destroying holes and (2) local antiferromagnetic recovery, each with corresponding rate.

Typical for $p$-doped copper oxides, the phase diagram in Fig. 3a shows that the N\'{e}el
temperature $T_{N}$---a measure for the stability of the antiferromagnetic phase---drops drastically with increased doping. At the doping rate $x = 0.02 \equiv x_{10}$ the crystal's repair effort loses out to the holes' destruction of loop currents. Thus antiferromagnetism disappears. Note that the doping rate $x_{10}$ corresponds, by Eqs. (1) and (2),  to a concentration of two holes per $10a_{0} \times 10a_{0}$ domain of the doping-pair superlattice. With more doping the \emph{additional} holes form a standing wave whose wave length equals the spacing of a ``free-hole'' superlattice,
\begin{equation}
\tilde{A}(\tilde{p}) = A(x - {x_{10}}) = \sqrt {\frac{2}{{x - {x_{10}}}}} \; a_{0} \; ,
\end{equation}
\noindent while a localized hole concentration $\hat{p}=x_{10}$ keeps suppressing the loop currents of antiferromagnetism from recurring.

\pagebreak
\includegraphics[width=5.95 in]{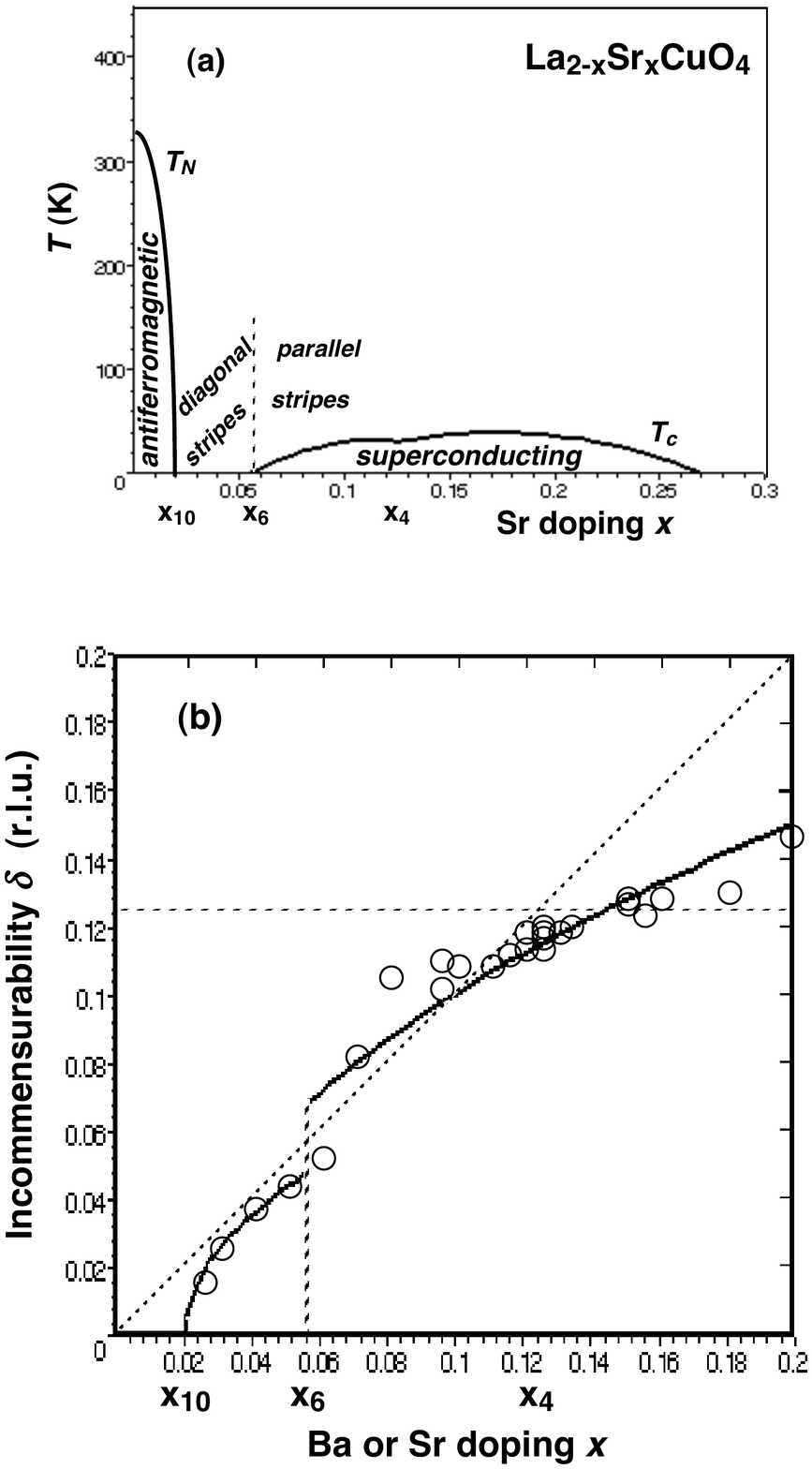}

\noindent FIG. 3. (a) Phase diagram of $p$-doped $La_{2-x}Sr_{x}CuO_{4}$.
(b) Incommensurability $\delta$ of magnetic density waves with wave number ${q_m} = \delta (2\pi /{a_0})$ and of charge-density waves with ${q_c} = 2\delta (2\pi /{a_0})$ in $La_{2-x}Ae_{x}CuO_{4}$ and in $La_{1.6-x}Nd_{0.4}Ae_{x}CuO_{4}$ due to doping with $Ae = Ba$ or $Sr$. Circles show data from neutron scattering or X-ray diffraction. The solid curve is a graph of Eq. (4). Dashed horizontal line at $\delta = 1/8$.
\pagebreak

The standing free-hole density-wave modulates the positions of the $Cu^{2+}$ and $O^{2-}$ ions in the $CuO_{2}$ plane by a periodic displacement. The resulting incommensurate charge pattern gives rise to satellite peaks in $X$-ray diffraction at reciprocal-lattice positions ($h \pm 2\delta ,k ,\ell  + \frac{1}{2}$) and ($h, k \pm 2\delta ,\ell  + \frac{1}{2}$). Here $h,k,\ell$ are integers, denoting prominent $X$-ray peaks. The term $\frac{1}{2}$ arises from an alternation of the planar hole-wave density distribution in successive unit cells along the $c$-axis.

The charge-incommensurability $2 \delta$ is given by the reciprocal spacing of the modulating free-hole superlattice, 
$2\delta (x) \equiv ({\Omega ^ \pm }/\sqrt 2 )/{\tilde{A}(\tilde{p})}$,
here more conveniently denoted, in $1/a_{0}$ units, as
\begin{equation}
\delta (x) = {\rm{ }}\frac{{{\Omega ^ \pm }}}{4}\sqrt {x - {x_{10}}} \;,
\end{equation}
\noindent with a stripe-orientation factor $\Omega^{+} = \sqrt 2$ for $x > x_{6}  \simeq 0.056$ when stripes are parallel to the $a$ or $b$ axis, or (derived for diagonal domains) $\Omega^{-} = 1$ for $x < x_{6}$ when stripes are diagonal.

The free-hole density wave not only displaces $Cu^{2+}$ and $O^{2-}$ ions but also
affects the ordering of their magnetic moments, causing satellite peaks to the antiferromagnetic Bragg peak, $\textbf{g} = (\frac{1}{2} \pm \delta ,\frac{1}{2} \pm \delta ,0)$, with the same incommensurability $\delta$, Eq. (4). Figure 3b shows a graph of the incommensurability (solid line) with data (circles) from both X-ray diffraction by charge-density waves and neutron scattering by magnetic density waves.\cite{4,5,6,7,8,9,10,11,12,13} It is observed that the density waves form stripes diagonal to the crystal axes in the very underdoped regime, $0.02  = x_{10}  < x < x_{6} \simeq 0.056$, extending the diagonal orientation of the destroyed antiferromagnetic phase up to the onset of superconductivity.\cite{2}  For higher doping, $x > x_{6}$, the orientation of the emerging superconducting electron pathways parallel to the crystal axes\cite{2} dominates and reorients the density waves.  The influence of the stripe orientation on the incommensurability $\delta$ is accounted for in Eq. (4) by the two-valued orientational factor $\Omega^{\pm}$, switching at $x = x_{6}$ with the onset of superconductivity.

The data in Fig. 3b fall close to the broken square-root graph of Eq. (4).  They skirt the diagonal line, $\delta = x$, in the underdoped regime, $x < x_{4} = 1/8$, as was first assumed to hold. However, near doping $x = x_{4}$, which is peculiar to superconductivity,\cite{2} there is a clear deviation, $\delta(x_{4}) < x_{4}$. The steady increase of $\delta$ in the overdoped regime,  $x > x_{4}$, shows that the incommensurability $\delta$ is determined only by the standing free-hole density wave in each $CuO_{2}$ plane and not by other circumstances (second ion-pair superlattice, staggered layer sandwiches) that affect superconductivity of $La_{2-x}Ae_{x}CuO_{4}$, $Ae = Sr$, $Ba$. 
The outlier, $\delta(x = 0.06)$, extending into the $x > x_{6}$ range, may result from orthorhombic distortion.\cite{2}

\pagebreak
\centerline{ \textbf{ACKNOWLEDGMENTS}}

\noindent I thank Duane Siemens for stimulating discussions and Preston Jones for help with LaTeX. Thanks also to Giacomo Ghiringhelli for providing an important literature reference.

\end{document}